\documentclass[a4paper,fleqn]{cas-dc}

\usepackage[authoryear,longnamesfirst]{natbib}

\def\tsc#1{\csdef{#1}{\textsc{\lowercase{#1}}\xspace}}
\tsc{WGM}
\tsc{QE}

\usepackage{amsmath}
\usepackage{amsfonts}				
\usepackage{mathrsfs} 				

\usepackage{bm}					
\usepackage{textcomp}				

\usepackage[hang,small,bf]{caption}
\usepackage[subrefformat=parens]{subcaption}
\usepackage{wrapfig}

\usepackage{color}

\usepackage{caption}
\usepackage{subcaption}

\usepackage{adjustbox}	

\usepackage{enumitem}		

\begin{document}
\let\WriteBookmarks\relax
\def\floatpagepagefraction{1}
\def\textpagefraction{.001}

\shorttitle{Spatially-Extended SpeF–Phixer for Spatial Causal Gene Inference}    

\shortauthors{T. Nagasaka}  

\title [mode = title]{Spatially-extended Flow Phixer (SpeF-Phixer): A Spatially Extended $\varphi$-Mixing Framework for Gene Regulatory Causal Inference in Spatial Gene Field


}  




%

\author[1,2]{Toru Nagasaka}[orcid=0009-0002-4322-4023]

\cormark[1]


\ead{toru-ngy@umin.ac.jp}


\credit{Conceptualization, Methodology, Software, Data curation, Visualization, Validation, Writing – review and editing}

\affiliation[1]{organization={Association of Medical Artificial Intelligence Curation},
            addressline={505, Sakae Members Office Building, 4-16-8 Sakae, Naka-ku}, 
            city={Nagoya},
            citysep={}, 
            postcode={460-0008}, 
            state={Aichi},
            country={Japan}}

\author[2]{Takaaki Tachibana}

\credit{Investigation, Data curation}

\affiliation[2]{organization={Division of Gastrointestinal Surgery, Department of Surgery,
Kobe University Graduate School of Medicine},
            addressline={7-5-2 Kusunoki-cho, Chuo-ku,}, 
            city={Kobe},
            citysep={}, 
            postcode={650-0017}, 
            state={Hyogo},
            country={Japan}}

\author[2]{Yukari Adachi}
\credit{Investigation, Data curation}

\author[2]{Hiroki Kagiyama}
\credit{Investigation, Data curation}

\author[2]{Ryota Ito}
\credit{Investigation, Data curation}

\author[3]{Mitsugu Fujita}


\credit{Writing – review and editing}

\affiliation[3]{organization={Center for Medical Education and Clinical Training, Kindai University Faculty of Medicine},
            addressline={1-14-1 Miharadai, Minami-ku}, 
            city={Sakai},
            citysep={}, 
            postcode={590-0111}, 
            state={Osaka},
            country={Japan}}

\author[4,2]{Kimihiro Yamashita}

\affiliation[4]{organization={Department of Biophysics, Kobe University Graduate School of Health Sciences},
            addressline={7-10-2 Tomogaoka, Suma-ku}, 
            city={Kobe},
            citysep={}, 
            postcode={654-0142}, 
            state={Hyogo},
            country={Japan}}

\credit{Resources, Project administration}

\author[2]{Yoshihiro Kakeji}

\credit{Supervision}

\cortext[1]{Corresponding author}

\fntext[1]{}


\begin{abstract}
\relax
\textbf{Background and objective:} Spatial transcriptomics provides rich spatial context but lacks sufficient resolution for large-scale causal inference. We developed SpeF--Phixer, a spatially extended $\phi$-mixing framework integrating whole-slide image (WSI)–derived spatial cell distributions with mapped scRNA-seq expression fields to infer directed gene regulatory triplets with spatial coherence.
\textbf{Methods:} Using CD103/CD8-immunostained colorectal cancer WSIs and publicly available scRNA-seq datasets, spatial gene fields were constructed around mapped cells and discretized for signed $\phi$-mixing computation. Pairwise dependencies, directional signs, and triplet structures were evaluated through $k$NN-based neighborhood screening and bootstrap consensus inference. Mediation and convergence were distinguished using generalized additive models (GAMs), with spatial validity assessed by real--null comparisons and database-backed direction checks.
\textbf{Results:} Across tissue patches, the pipeline reduced $\sim\!3.6\times10^4$ triplet candidates to a reproducible consensus set ($\sim\!3\times10^2$ per patch). The downstream edge ($Y\!\rightarrow Z$) showed significant directional bias consistent with curated regulatory databases. Spatial path tracing demonstrated markedly higher coherence for real triplets than for null controls, indicating that inferred chains represent biologically instantiated regulatory flows.
\textbf{Conclusion:} SpeF--Phixer extracts spatially coherent, directionally consistent gene regulatory triplets from histological images. This framework bridges single-cell molecular profiles with microenvironmental organization and provides a scalable foundation for constructing spatially informed causal gene networks.
\end{abstract}





\maketitle


\section{Introduction}

Inferring gene regulatory networks (GRNs) from transcriptomic data is a
central task in systems biology, yet remains fundamentally challenging.
Classical approaches---including correlation-based clustering, regression and
tree-ensemble prediction, Bayesian and dynamic Bayesian networks, and
information-theoretic methods such as ARACNE \citep{margolin2006aracne}, CLR \citep{faith2007large}, MRNET \citep{meyer2007information}, and PIDC \citep{chan2017gene}---estimate
pairwise relationships from bulk or single-cell RNA expression profiles
\citep{saint2020network}. These families differ in their statistical
assumptions and computational demands, but they share a key limitation:
most operate on expression matrices alone and therefore cannot distinguish
direct from indirect interactions, causal directionality, or context-specific
regulatory programs such as those dictated by tissue architecture or
cell–cell proximity.

Among existing approaches, Phixer \citep{singh2019inferring} is unique in its
use of the $\phi$-mixing coefficient, a directional and asymmetric dependency
measure that naturally encodes $X_i\to X_j$ versus $X_j\to X_i$, and permits
cycles—properties well aligned with biological regulation. Phixer further uses
a triplet-level pruning rule reminiscent of the Data Processing Inequality
to suppress cascade-type false positives. However, $\phi$-mixing inference
remains constrained by the same limitation shared by other GRN methods: it
treats each cell as an isolated expression vector and ignores the spatial
organization and phenotypic heterogeneity that shape gene–gene dependencies
in real tissues.

Recent developments in spatial transcriptomics deconvolution
(e.g.\ CIBERSORTx \citep{newman2019determining}, SPOTlight \citep{elosua2021spotlight}, RCTD/Spacexr \citep{cable2022robust}, Cell2location \citep{kleshchevnikov2022cell2location}, Tangram \citep{biancalani2021deep})
illustrate the transformative impact of integrating histological context,
but these frameworks address tissue composition rather than causal
gene–gene relationships. Likewise, deep-learning–based cell classification
models accurately identify cell types within whole-slide images (WSI) but have
not been systematically integrated with GRN inference. Thus, current GRN
methods lack a mechanism to incorporate \emph{where} a cell is located, what
phenotypes surround it, and how local tissue microenvironments modulate
molecular interactions.

To overcome these limitations, we introduce \textbf{SpeF-Phixer}
(Spatially-extended Flow Phixer), a spatially aware extension of the original
Phixer algorithm that incorporates (i) high-resolution tissue architecture
via AI-based cell detection and classification and (ii) spatially stratified
single-cell expression profiles mapped onto histological coordinates.
Instead of inferring edges solely from bulk or scRNA-seq distributions,
SpeF-Phixer estimates \emph{triplet-level directional structure} 
$X\to Y\to Z$ by combining $\phi$-mixing asymmetry with 
tissue-resolved co-localization, neighborhood composition, and
cell–cell interaction patterns. This spatial augmentation resolves
ambiguities intrinsic to expression-only inference, particularly for
triplet configurations that are topologically indistinguishable in
non-spatial GRN models.

In this study, we leverage AI-guided cellular phenotyping from whole-slide
immunostained colorectal cancer tissues and map these image-derived cell types
to scRNA-seq expression profiles. By constructing \emph{spatially grounded}
directional $\phi$-mixing matrices and performing consensus estimation across
local microenvironments, SpeF-Phixer yields more accurate and biologically
coherent triplet predictions than expression-only methods. This framework
bridges digital pathology and causal network inference, enabling the
identification of regulatory motifs that specifically operate at tumor–immune
interfaces, stromal niches, and tissue-resident memory T cell (Trm)–rich
microenvironments—contexts that are invisible to conventional GRN algorithms.

\section{Materials}

\subsection{Patients and Samples}

This study included tissue specimens from 303 colorectal cancer patients who underwent biopsy or surgery, with or without neoadjuvant chemoradiotherapy. Both biopsy and surgical samples were formalin-fixed and paraffin-embedded, and were subjected to dual immunostaining for CD103 and CD8. The staining process was performed according to established protocols to ensure consistency and reliability \citep{ohno2024tumor}.

\begin{figure}[h]
\begin{minipage}[t]{0.48\linewidth}
\includegraphics[width=1.0\columnwidth]{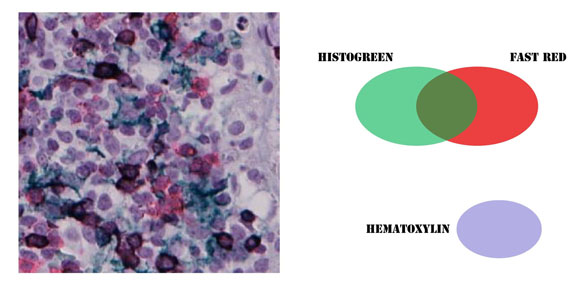}
\centering 
\end{minipage}
\begin{minipage}[t]{0.48\linewidth}
\includegraphics[width=1.0\columnwidth]{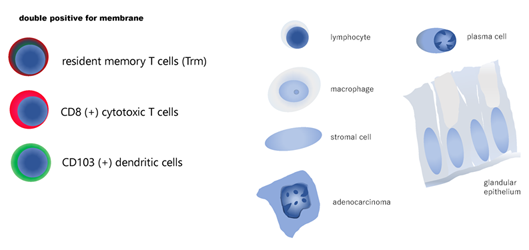}
\centering 
\end{minipage}

\caption{CD103/CD8 double stain}
\label{fig:CD103_CD8}
\end{figure}

\subsection{Data Annotation Methodology Using Cu-Cyto\textsuperscript{\tiny{\textregistered}} Viewer}
Cell-level annotation was performed using the Cu-Cyto\textsuperscript{\tiny{\textregistered}} Viewer, a specialized annotation tool designed for precise cellular identification in histopathological images \citep{abe2023deep}. The annotation workflow employed a human-AI collaborative approach to ensure comprehensive cellular coverage and annotation accuracy across all tissue sections.

\begin{figure}[htbp]
\centering
\begin{subfigure}[b]{0.48\columnwidth}
    \centering
    \includegraphics[width=\columnwidth]{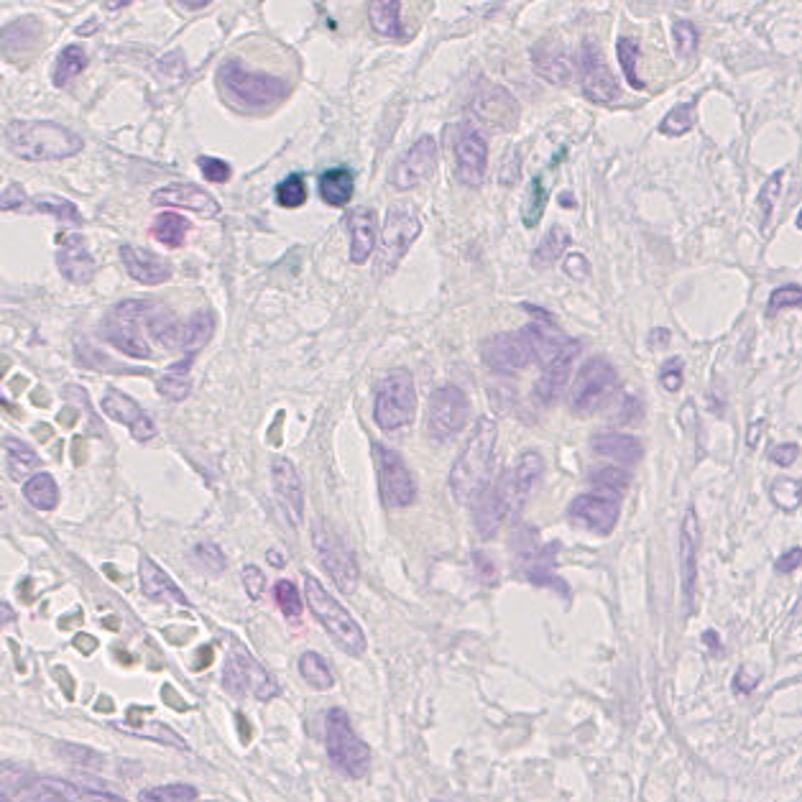}
    \caption{Original tissue section without annotations.  
    Neoplastic epithelial cells form infiltrative small nests within the stromal tissue, illustrating characteristic patterns of colorectal adenocarcinoma invasion.}
    \label{fig:before_annotation}
\end{subfigure}
\hfill
\begin{subfigure}[b]{0.48\columnwidth}
    \centering
    \includegraphics[width=\columnwidth]{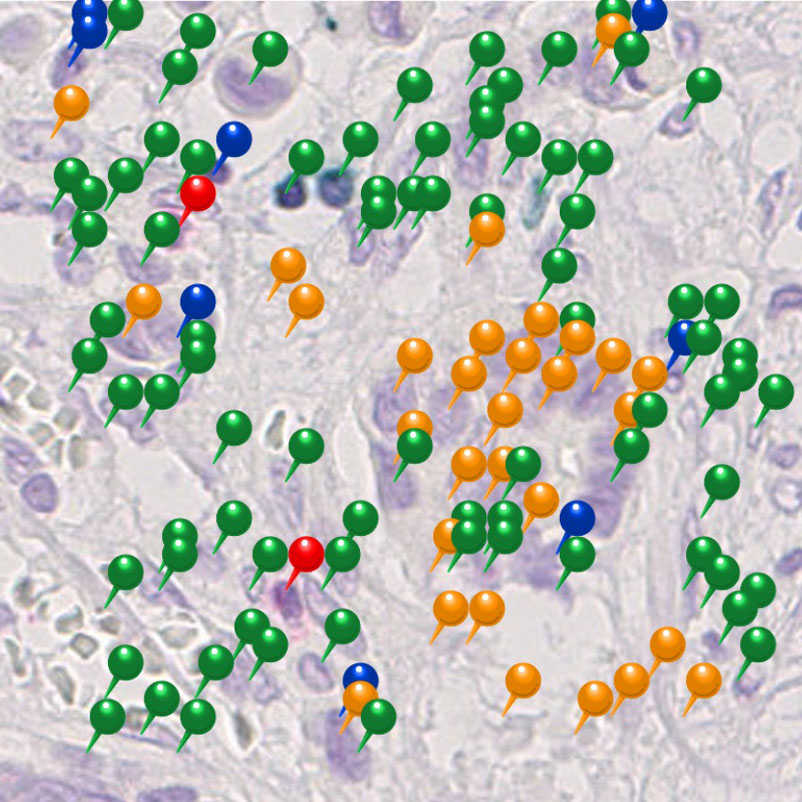}
    \caption{Cellular annotations using colored markers:  
    orange pins indicate cancer cells, red pins represent CD8(+) T lymphocytes, blue pins correspond to lymphocytes, and green pins denote other cell populations.}
    \label{fig:after_annotation}
\end{subfigure}
\caption{Cu-Cyto\textsuperscript{\tiny{\textregistered}} Viewer annotation workflow demonstration. (a) Unannotated histological section showing invasive carcinoma architecture. (b) The same section annotated with color-coded cellular markers for spatial mapping and quantitative analysis.}
\label{fig:annotation_workflow}
\end{figure}

The annotation process consisted of a two-stage protocol (Figure \ref{fig:annotation_workflow}). Initially, a prototype AI model performed automated cell detection and placed preliminary annotation markers across the tissue sections. Subsequently, expert annotators systematically reviewed these AI-generated annotations using the Cu-Cyto\textsuperscript{\tiny{\textregistered}} Viewer interface, performing two primary tasks: (1) verification of AI-placed markers for accuracy, (2) addition of markers for cells missed by the automated detection system.

The annotation system employed a color-coded marking scheme where distinct colors represented different cellular populations and morphological categories. This visual encoding facilitated rapid identification of annotation patterns and enabled efficient quality control assessment across multiple annotators.
Annotation quality was maintained through a multi-stage verification protocol. Following the initial annotation refinement by expert annotators, a second expert annotator performed a double-check of the annotations, with a final review conducted by a pathologist to ensure the highest level of accuracy and reliability. Inter-annotator agreement was assessed through systematic comparison of overlapping annotation regions, ensuring consistent application of classification criteria across all tissue sections.
The Cu-Cyto\textsuperscript{\tiny{\textregistered}} Viewer platform recorded detailed annotation metadata including marker coordinates, classification confidence levels, and annotator identification for traceability. This comprehensive documentation enabled retrospective quality assessment and facilitated standardized patch extraction procedures for subsequent model training and validation processes.

This protocol required complete cellular coverage, ensuring that every identifiable cell within the tissue section received appropriate classification markers, which serves as the basis for calculating the estimated true prior probabilities for cell classification.

\subsection{AI-based Cell Classification and Training Dataset}

Whole-slide images (WSI) immunostained for CD103 and CD8 were processed
using a human–AI collaborative annotation workflow.
Instead of automated nuclear detection by color deconvolution and
watershed segmentation, cell-centered image patches (40\,$\times$\,40 pixels)
were directly extracted around expert-annotated marker coordinates generated
in the Cu-Cyto\textsuperscript{\tiny{\textregistered}} Viewer.
This ensured that the training data relied on verified cellular identities
rather than algorithmic heuristics.

To construct balanced and statistically reliable training data, we applied a
sampling threshold (\texttt{flag\_limit}) of 11{,}000 patches per cell class.
When a class exceeded this threshold, patches were randomly subsampled;
when below, all available patches were included.

Two neural network architectures were employed for supervised cell-type
classification: (i) a residual network (ResNet) \citep{he2016deep}, and (ii) a customized
Vision Transformer (ViT) \citep{dosovitskiy2021imageworth16x16words} optimized for small patch embeddings.
For each cell class, both models were trained under identical conditions,
and the model achieving the higher validation performance was used in
downstream inference.
In practice, the ViT implementation yielded superior feature separability
in the majority of cell classes and was therefore selected for most WSIs.

In our previous work \citep{tachibana2025reliability}, which analyzed
185{,}432 annotated cell images across 16 cell types, feature separability in
the learned embedding space was quantitatively evaluated using
cosine-similarity–based likelihood quality scores.
That framework demonstrated that the resulting feature representations are
reliable and robust to prior probability shift, establishing the
trustworthiness of the learned embeddings for downstream biological analysis.

The trained deep-learning models were subsequently applied to all WSIs in the
present study to generate per-cell phenotype labels.
These image-derived cell types were used solely as inputs for spatial
integration in the SpeF--Phixer workflow, and the detailed AI methodology
follows the validated approach of our previous reliability assessment study
\citep{tachibana2025reliability}.

\subsection{Single-cell RNA sequencing dataset and preprocessing}

Publicly available single-cell RNA sequencing (scRNA-seq) data of human colorectal cancer \citep{lee2024comprehensive} were obtained from the Gene Expression Omnibus (GEO).  
We used the dataset \textbf{GSE132465}, which provides 10x Genomics–based single-cell transcriptomes with accompanying cell-type annotations.
Only processed matrices supplied by the authors were used, including:
\begin{itemize}[leftmargin=0pt, itemindent=5pt, labelsep=5pt]
    \item \url{GSE132465_GEO_processed_CRC_10X_cell_annotation.txt.gz}
    \item \url{GSE132465_GEO_processed_CRC_10X_natural_log_TPM_matrix.txt.gz}
    \item \url{GSE132465_GEO_processed_CRC_10X_raw_UMI_count_matrix.txt.gz}
\end{itemize}

Raw UMI counts were used exclusively for gene-level statistical evaluation (mean, variance, coefficient of variation, percentage of expressing cells), while log-TPM matrices were used for downstream gene panel construction, cell-type scoring, and creation of expression reference profiles.

The matrices were downloaded programmatically and converted into an \texttt{AnnData} object using the \texttt{scanpy} framework. Cells with missing annotations or mismatched barcodes were discarded. All analyses were performed using custom Python code.

\subsection{Gene panel selection strategy}

To create a biologically meaningful and computationally efficient gene panel linking scRNA-seq profiles with image-derived cell classes, we combined data-driven highly variable gene (HVG) selection with expert-curated lineage markers relevant to colorectal cancer biology.

\subsubsection*{(1) Highly variable genes (HVGs)}

To quantify dispersion relative to mean expression, we used the coefficient of
variation (CV). For each gene in the raw UMI matrix, we computed:
\[
\text{CV} = \frac{\sqrt{\mathrm{Var}(g)}}{\mathrm{Mean}(g) + 10^{-10}}.
\]

Genes were filtered based on the percentage of non-zero expression:
\[
1\% \le \text{pct.cell.expr} \le 99\%.
\]

From the filtered pool, the top 2,000 genes ranked by CV were selected as HVGs.

\subsubsection*{(2) Curated cell-type marker genes}

Curated marker genes were compiled for major epithelial, immune, stromal, and tumor-associated lineages relevant to colorectal cancer microenvironmental analysis.  
Representative examples are shown below.

\begin{table}[htbp]
\centering
\footnotesize
\adjustbox{width=\columnwidth}{		
\begin{tabular}{ll}
\hline
\textbf{Lineage} & \textbf{Representative markers} \\
\hline
Epithelial & EPCAM, KRT8, KRT18, CEACAM5 \\
Stem-like & LGR5, ASCL2, OLFM4 \\
CD8 T cells & CD8A, CD8B, GZMB, PRF1 \\
Tissue-resident memory T (Trm) & ITGAE (CD103), CD69, CXCR6, ZNF683 \\
NK cells & KLRD1, NKG7, GNLY \\
B cells / Plasma cells & MS4A1, CD79A, JCHAIN, MZB1 \\
Myeloid / Macrophage & LYZ, CD68, CD163, MRC1 \\
Fibroblasts / CAFs & COL1A1, PDGFRA, FAP, ACTA2 \\
Endothelial & PECAM1, CDH5, VWF \\
Adenocarcinoma markers & EPCAM, KRT19, MKI67, ETV4, MMP9 \\
\hline
\end{tabular}
}	
\caption{Curated marker genes used for cell-type identification and Trm profiling.}
\end{table}

All gene symbols were harmonized using uppercase HGNC names, with synonomy mapping (e.g., ITGAE = CD103; PECAM1 = CD31).

\subsubsection*{(3) Hybrid gene panel construction}

We constructed multiple candidate gene panels and used the panel optimized for downstream expression mapping:

\[
\text{Gene panel} = \text{HVGs} \cup \text{Existing curated markers}.
\]

A more focused panel combining the top 1,000 HVGs with all validated lineage markers (Trm, epithelial, stromal, and immune families) was ultimately selected, yielding:
\[
\sim 1,500\text{--}2,000\ \text{genes (dataset-dependent)}.
\]

This hybrid panel preserves expressive variation while ensuring physiologically interpretable lineage discrimination.

\subsubsection*{(4) Marker-rule–based cell-type identification}

For mapping scRNA-seq cell states to image-derived AI classes, each AI label was linked to a rule-based marker signature of the form:
\[
\text{all\_of (AND)},\quad \text{any\_of (OR)},\quad \text{not\_any (NOT)},
\]
with each clause evaluated using quantile thresholds of non-zero expression.  
For example, CD8 Trm cells were identified using:

\[
\begin{aligned}
\text{CD8A}_{q>0.5},\ \text{CD8B}_{q>0.5},\ \text{ITGAE (CD103)}_{q>0.5},\\
\text{and optional }\text{CD69}_{q>0.5}.
\end{aligned}
\]

Cells matching a rule were aggregated to compute mean and variance vectors, producing reference expression profiles for each AI class.

These profiles were later used to assign gene-level expression to image-derived cells during multimodal integration.

\subsubsection*{(5) Selection of spatial patches for SpeF--Phixer analysis}

For downstream spatial causal inference, we extracted fixed-size image
patches of $802 \times 802$ pixels (corresponding to $146.766\,\mu\mathrm{m}$
at a resolution of $0.183\,\mu\mathrm{m}$/pixel).  
A total of 25 patches (5 cases $\times$ 5 patches per case) were selected
from regions containing both CD103$^{+}$ and CD8$^{+}$ immune cells together
with cancer cell nests.  
To avoid information leakage, all patches were taken from anatomical
locations distinct from those used for training the Cu-Cyto\textsuperscript{\tiny\textregistered}
AI models.  
Patches meeting these criteria were then sampled at random for SpeF--Phixer
processing, ensuring unbiased coverage of tumor–immune interface regions
across specimens.

\section{Methods}

\subsection{Spatially-extended Flow Phixer (SpeF--Phixer) algorithm}

\subsubsection*{Algorithm overview}

The Spatially-extended Flow Phixer (SpeF--Phixer) algorithm extends the
$\phi$-mixing--based causal inference framework of Phixer \citep{singh2019inferring}
to spatially resolved histological data. Using (i) cell-level class probability
fields obtained from whole-slide image (WSI) AI models and (ii) gene expression
fields mapped from single-cell RNA sequencing (scRNA-seq), SpeF--Phixer estimates
directed dependencies between genes or cell states while preserving spatial
context. Unlike correlation- or regression-based approaches, $\phi$-mixing
captures conditional directional dependence and allows asymmetric inference
$A\!\rightarrow\!B \neq B\!\rightarrow\!A$ under non-Gaussian, non-linear
conditions.

As inputs, we use gene expression profiles from sc/snRNA-seq data and spatial
coordinates and cell-type probabilities obtained from WSI AI analysis.
scRNA-seq--derived cell embeddings are aligned to histological coordinates by
probabilistic matching of cell-type distributions, morphological features, and
local gene expression similarities, yielding a one-to-one correspondence between
expression vectors and spatial positions.

The extraction of spatially stable regulatory triplets proceeds through a
multi-stage screening pipeline that progressively reduces the search space from
the full marker panel to a reproducible subset of directionally coherent
three-gene chains. Screening is first conducted at the pairwise level to
identify spatially co-varying candidate edges, which are then scored by
$\phi$-mixing and assembled into directed triplets. The overall algorithm
therefore proceeds through: (1) spatial and molecular field construction,
(2) neighborhood-based pairwise gene screening,
(3) signed $\phi$-mixing computation,
(4) bootstrap-based consensus inference of directed edges and triplets,
(5) mediation–convergence discrimination for consensus triplets,
 and (6) spatial visualization.

\subsubsection*{Construction of the spatial gene field}
For each cell $i$ at spatial location $(x_i,y_i)$ with normalized expression 
$g_i$, a spatially diffused expression potential $g_i'$ is computed based on 
the spatial interaction potential framework \citep{nagasaka2024novel} as
\[
  g_i' = \sum_{j \ne i} \frac{g_j}{d_{ij}^2 + \epsilon},
  \qquad
  d_{ij}^2 = (x_i - x_j)^2 + (y_i - y_j)^2.
\]
The spatial coordinates $(x_j, y_j)$ are assigned probabilistically based on 
the softmax output $p_j$ of the trained classifier, which represents the 
probability of each cell belonging to its assigned cell type 
\citep{tachibana2025reliability}. The uncertainty associated with this 
probabilistic cell-type assignment is absorbed into the bootstrap resampling 
procedure (see Bootstrap and consensus estimation), which evaluates the 
robustness of inferred dependencies across spatially resampled configurations.
This inverse-square--weighted aggregation yields continuous spatial gene fields 
reflecting the influence of nearby cells while downweighting distant contributions.  
A small constant $\epsilon$ ensures numerical stability, and values may 
optionally be log-transformed as $\log(1+g_i')$ to suppress heavy-tailed 
distributions.  
The resulting variables (\texttt{gene\_A\_field}, \texttt{gene\_B\_field}, 
\texttt{gene\_C\_field}, \dots) serve as spatially extended inputs for 
$\phi$-mixing--based dependency inference.

\subsubsection*{Neighborhood-based pairwise gene screening}
To control the combinatorial explosion of triplets, SpeF--Phixer first identifies spatially co-varying gene pairs on each patch.  
For every anchor gene $A$, two complementary screening modules are applied to the mapped expression fields:

\begin{itemize}
\item \texttt{select\_topk\_neighbors\_by\_corr\_sparse}:  
detects positively co-varying neighbors by computing sparse local correlations across $k$-nearest neighbor (kNN) neighborhoods on the tissue graph.

\item \texttt{detect\_repressive\_relationships}:  
detects inhibitory neighbors by testing whether expression of $A$ reduces a candidate gene $B$ within cells in which $B$ is normally expressed, based on negative correlation and decreased conditional mean of $B$ when $A$ is on.
\end{itemize}

The two neighbor lists are merged to form, for each gene $A$, a combined set of promotive and repressive neighbors $(A,B)$ together with correlation- and suppression-based scores.  
This step provides a biologically informed pre-screening in which candidate promotive pairs are expected to correspond to positive signed $\phi$ values and repressive pairs to negative signed $\phi$ values, while discarding gene pairs that show little evidence of interaction in the local spatial context.  
Across the 1{,}155 expressed genes per patch, this yields approximately 
$3.8 \times 10^{3}$ directed neighbor pairs that enter the subsequent 
$\phi$-mixing analysis, giving rise to approximately $3.6 \times 10^{4}$ 
possible triplet combinations.

\subsubsection*{Discretization and $\phi$-mixing--based directed dependency inference}
Each spatially diffused gene field is discretized into $Q$ equal-width bins, with bin edges determined from the pooled 1--99th percentile range of all values to ensure a common discretization scheme across genes.

For discretized variables $(X_i, X_j)$, the $\phi$-mixing coefficient \citep{singh2019inferring} is defined in the discrete setting as
\[
  \phi(X_j \mid X_i)
  = \max_{b}
    \frac{1}{2}
    \sum_{a}
      \bigl|
        P(X_j = a \mid X_i = b) - P(X_j = a)
      \bigr|,
\]
which measures the maximal deviation between the marginal and $X_i$-conditioned distributions of $X_j$.  
Evaluating this coefficient for all ordered pairs yields an asymmetric dependency matrix
$\Phi = [\phi(X_j \mid X_i)]$.

Edges with $\phi(X_j \mid X_i)$ above a predefined threshold are retained as directed dependencies $X_i\!\to\!X_j$.  
Indirect or cascade-type edges are removed using the standard Phixer pruning rule, a discrete analogue of the Data Processing Inequality:
\[
  \phi(X_k \mid X_i) 
  \;\le\;
  \min\!\bigl( 
    \phi(X_j \mid X_i),\,
    \phi(X_k \mid X_j)
  \bigr),
\]
in which case the edge $X_i \to X_k$ is pruned.  
After pruning, the remaining edges define a directed dependency graph 
$G = (V, E, W)$ with edge weights $w_{ij} = \phi(X_j \mid X_i)$.

\subsubsection*{Directional sign from discrete conditional divergence}
While $\phi(X_j \mid X_i)$ quantifies asymmetric dependency strength, it does not indicate whether $X_i$ promotes or inhibits $X_j$.  
To assign directional polarity, SpeF--Phixer augments $\phi(X_j\mid X_i)$ with a sign computed directly from the discretized conditional distribution of $X_j$.

Given discretized variables $(X_i, X_j)$ taking values in 
$\{0,\dots,Q_i-1\}$ and $\{0,\dots,Q_j-1\}$, respectively, the total variation (TV) deviation of $X_j$ under each bin $b$ of $X_i$ is
\[
  \mathrm{TV}(X_j \mid X_i=b)
  =
  \tfrac{1}{2}\sum_{a=0}^{Q_j-1}
    \bigl|P(X_j=a\mid X_i=b) - P(X_j=a)\bigr|.
\]
The most perturbed bin is selected as
\[
  b^\ast = \arg\max_b \mathrm{TV}(X_j \mid X_i=b).
\]
This bin represents the condition under which $X_i$ induces the largest distributional shift in $X_j$ and thus provides the most informative basis for determining the direction of influence.

Let $\mu(X_j\mid X_i=b^\ast)$ denote the conditional mean of $X_j$
in this maximally perturbed bin and $\mu(X_j)$ the global mean.  
SpeF--Phixer assigns a directional label according to
\[
\mathrm{dir}_{i\to j} =
\begin{cases}
  \mathrm{promote}, & \mu(X_j\mid X_i=b^\ast) > \mu(X_j), \\[4pt]
  \mathrm{inhibit}, & \mu(X_j\mid X_i=b^\ast) < \mu(X_j), \\[4pt]
  \mathrm{neutral}, & \text{otherwise}.
\end{cases}
\]
A signed $\phi$-mixing coefficient is then constructed as
\[
  \phi^{\mathrm{signed}}_{j\mid i}
  = s_{i\to j}\,\phi(X_j\mid X_i),
  \qquad
  s_{i\to j} \in \{-1,0,+1\},
\]
where $s_{i\to j}=+1$ for promotive, $-1$ for inhibitory, and $0$ for neutral relations.  
This sign-refined coefficient allows SpeF--Phixer to distinguish 
promotive ($X_i$ high $\Rightarrow$ $X_j$ high) and inhibitory
($X_i$ high $\Rightarrow$ $X_j$ low) dependencies even when the global correlation between $X_i$ and $X_j$ is weak or misleading.

\subsubsection*{Triplet inference from the signed $\Phi$ matrix}
Given the signed matrix
\[
  \Phi^{\mathrm{signed}} = 
  \bigl[\phi^{\mathrm{signed}}_{j\mid i}\bigr]_{i,j=1}^n,
\]
each entry $\phi^{\mathrm{signed}}_{j\mid i}$ encodes the directed dependence from $i$ to $j$, combining both magnitude and polarity (promotive $+$ or inhibitory $-$).  
This asymmetric matrix defines a weighted directed graph $G=(V,E,W)$ with edge weights $w_{ij}=\phi^{\mathrm{signed}}_{j\mid i}$.

For each unordered triplet of genes $\{A,B,C\}$, SpeF--Phixer evaluates all six directional permutations
$(a,b,c)\in\mathrm{Perm}(A,B,C)$
and selects the most coherent causal chain $a\!\to\!b\!\to\!c$ under a sign-consistency constraint.  
Let
\[
s_{b\mid a}=\mathrm{sign}\!\bigl(\phi^{\mathrm{signed}}_{b\mid a}\bigr),\qquad
s_{c\mid b}=\mathrm{sign}\!\bigl(\phi^{\mathrm{signed}}_{c\mid b}\bigr),
\]
and define the chain strength and the sign-coherence indicator as
\[
m(a,b,c)=\min\!\bigl(\,|\phi^{\mathrm{signed}}_{b\mid a}|,\;|\phi^{\mathrm{signed}}_{c\mid b}|\,\bigr),\qquad \\
\chi(a,b,c)=\mathbb{I}\{\,s_{b\mid a}=s_{c\mid b}\,\}.
\]
The triplet score is then
\[
S(a,b,c)=m(a,b,c)\cdot \chi(a,b,c),
\]
which requires both consecutive links to be strong and of the same polarity (promotive--promotive or inhibitory--inhibitory).

The optimal permutation and its score are
\[
\hat{\pi}=\underset{(a,b,c)\in\mathrm{Perm}(A,B,C)}{\arg\max}\; S(a,b,c),
\qquad
\hat{S}=S\bigl(\hat{\pi}\bigr).
\]
Let $(X,Y,Z)=\hat{\pi}$ denote the canonical ordering for this triplet.
The triplet polarity (promotive or inhibitory) is inherited from the edge sign
$s_{Y\mid X}(=s_{Z\mid Y})$, and the signed edge coefficients
$\phi^{\mathrm{signed}}_{Y\mid X}$ and $\phi^{\mathrm{signed}}_{Z\mid Y}$ 
are recorded together with $\hat{S}$.  
Triplets with $\hat{S}$ below a predefined path-score threshold are discarded.

This canonicalization $(X\!\to\!Y\!\to\!Z)$ is stored together with a mapping to the original labels $(A,B,C)$
to ensure traceability. Parallel dependencies such as $A\!\to\!C$ or $B\!\to\!C$
are implicitly evaluated through alternative permutations during the maximization,
while indirect edges have already been pruned at the pairwise inference stage.

\subsubsection*{Bootstrap and consensus estimation}
To assess robustness against sampling variance, the algorithm performs $B$ bootstrap iterations. 
In each iteration, subsets of spatial points and gene fields are resampled with replacement, a signed $\Phi^{\mathrm{signed}}$ matrix is recomputed, and the set of best-scoring triplets $\mathcal{T}^{(b)}=\{(X,Y,Z)_b\}$ is collected.

Consensus statistics are then derived across bootstraps:
\begin{align}
\mathrm{count}(X,Y,Z) &= \sum_{b=1}^{B} \mathbb{I}\{(X,Y,Z)\in \mathcal{T}^{(b)}\}, \\
\pi(X,Y,Z) &= \frac{\mathrm{count}(X,Y,Z)}{B}, \\
\bar{\phi}_{Y\mid X} &= \frac{1}{B}\sum_{b}\phi^{\pm}_b(Y\mid X), \\
\bar{\phi}_{Z\mid Y} &= \frac{1}{B}\sum_{b}\phi^{\pm}_b(Z\mid Y).
\end{align}

Triplets satisfying $\pi(X,Y,Z)>\tau_{\mathrm{cons}}$ (typically $\tau_{\mathrm{cons}}=0.6$) are regarded as consensus causal chains.  
The direction consistency ratio, i.e., the fraction of bootstraps with concordant promotive/inhibitory sign, is recorded as the consensus direction.  
Edges $(Y,Z)$ are simultaneously aggregated across all bootstraps to form a consensus edge list with analogous statistics.

\subsubsection*{Directional normalization and canonical output}
All retained triplets are reported in a direction-normalized form $(X\rightarrow Y\rightarrow Z)$ 
together with the corresponding signed coefficients $\bar{\phi}^{\pm}(Y\mid X)$ and $\bar{\phi}^{\pm}(Z\mid Y)$.  
The canonical and raw indices $(A,B,C)$ are both stored in separate mapping tables to ensure traceability of the permutation applied during optimization.  
Consensus edges and triplets are then exported as tabular summaries 
(\texttt{consensus\_edges.csv}, \texttt{consensus\_triplets.csv}) 
for downstream graph assembly and spatial visualization.

\subsubsection*{Mediation versus convergence discrimination}
To distinguish mediated chains ($X \rightarrow Y \rightarrow Z$) from convergent inputs 
($X \rightarrow Z$, $Y \rightarrow Z$), 
conditional and ensemble statistics were computed on the consensus triplets.
For each retained triplet $(X,Y,Z)$, gene-wise expression profiles along pseudotime $\tau$
were modeled by generalized additive models (GAMs) with smooth splines 
$s(\cdot)$ while controlling for $\tau$:
\begin{align}
\text{Model}_1: &\quad Z \sim s(\tau) + s(X), \\
\text{Model}_2: &\quad Z \sim s(\tau) + s(Y), \\
\text{Model}_3: &\quad Z \sim s(\tau) + s(X) + s(Y).
\end{align}
A chain was regarded as \emph{mediated} when the inclusion of $Y$ in Model$_3$ 
led to a substantial reduction of the partial effect of $X$
($p_X > 0.05$ or $\Delta\mathrm{AIC}_{X} > 0$), 
together with a significant contribution of $Y$ ($p_Y < 0.05$).
Conversely, chains were classified as \emph{convergent}
when both $s(X)$ and $s(Y)$ remained independently significant in Model$_3$ 
and removal of either term markedly worsened the fit 
($\Delta\mathrm{AIC} > 2$ or $\Delta\mathrm{GCV} > 0.01$).
To account for sampling variability, the above GAM fits were repeated in
$B$ block-bootstrap replicates preserving pseudotemporal ordering,
yielding bootstrap support probabilities for the mediation ($p_{\mathrm{med}}$)
and convergence ($p_{\mathrm{con}}$) hypotheses.

In parallel, partial $\Phi$-mixing coefficients were recomputed under
conditional formulations,
\[
\phi^{\pm}_{Z\mid X;Y},\quad \phi^{\pm}_{Z\mid Y;X},
\]
to quantify the residual directed dependence after controlling for the other mediator.
A triplet was labeled \emph{mediated} when
$|\phi_{Z\mid X;Y}| < 0.5|\phi_{Z\mid X}|$ and $p_{\mathrm{med}} > 0.6$,
\emph{convergent} when
both conditional coefficients remained large
($|\phi_{Z\mid X;Y}|, |\phi_{Z\mid Y;X}| > 0.5|\phi_{Z\mid X}|, |\phi_{Z\mid Y}|$)
and $p_{\mathrm{con}} > 0.6$, 
and \emph{ambiguous} otherwise.

The resulting mediation and convergence support values were appended to the
consensus triplet tables and propagated as edge-type annotations during network
assembly. In the subsequent graph-level pruning step, 
edges involved in strongly mediated chains were downweighted or removed 
(\emph{transitive reduction}), whereas convergent inputs were retained as
independent regulators of $Z$.

\subsubsection*{Spatial visualization of causal flows}
Two complementary visualization modules were implemented:
\begin{enumerate}
  \item \textbf{KDE Edge Bundling.}  
  Paths connecting spatially adjacent cells (via $k$-nearest neighbors) 
  are projected onto a 2D density map.  
  By iteratively shifting points toward the gradient ridges of this density, 
  trajectories are aggregated into smooth bundles representing 
  \textit{spatial pathway density}.  
  These ridges correspond to the most frequently traversed regions of 
  inferred causal flows.
  \item \textbf{Streamline Visualization.}  
  Gene expression fields $B(x,y)$ and $C(x,y)$ are smoothed using 
  Gaussian filtering, and their gradients $\nabla B$, $\nabla C$ are computed.
  Streamlines are drawn along these vector fields to depict the 
  \textit{direction of gene expression gradients}.
\end{enumerate}

\subsection{Statistical validation procedures}

\subsubsection*{Null triplets and path drawing rate}
To assess whether the spatial flow patterns observed for SpeF--Phixer triplets
could arise by chance under the same expression landscape, we constructed
a null ensemble of \emph{random triplets} on each eligible patch and
compared the corresponding ``path drawing rate'' against that of the
GAM-supported consensus triplets.

Eligible patches were defined as described above based on tumor content
and quality control criteria (tumor area fraction within the predefined
range, simultaneous presence of tumor and stroma, and absence of major
artifacts). For each eligible patch, we first assembled a \emph{gene pool}
from the mapped single-cell expression matrix $E$ restricted to that patch.
Genes were required to be present with the specified prefix (e.g.\ \texttt{gene\_}),
to exhibit nonzero expression in at least a minimal fraction of mapped cells,
and to show non-negligible variance so as to exclude genes that are essentially
absent or constant on the patch. The SpeF--Phixer consensus triplets $(X,Y,Z)$
with GAM support for mediation or convergence (i.e.\ classified as non-ambiguous
by the GAM-based procedure above) were then collected into the set
$\mathcal{T}_{\mathrm{real}}$ for that patch.

For each patch, we sampled without replacement $|\mathcal{T}_{\mathrm{real}}|$
\emph{null triplets} $(X',Y',Z')$ by drawing three distinct genes uniformly
from the gene pool, explicitly excluding any triplet already present in
$\mathcal{T}_{\mathrm{real}}$. Thus, the null triplets matched the real set in
cardinality and marginal expression characteristics (same mapped cells, same
gene pool, identical preprocessing), but were not constrained by
$\Phi$-mixing or GAM criteria.

For every real and null triplet, spatial bundles and streamlines were
generated using the same KDE edge-bundling and streamline visualization
modules, with identical $k$-nearest neighbor graphs, kernel parameters, and
rendering thresholds. A triplet was counted as \emph{successfully drawn} in a
given mode (mediated or converged) if the bundle overlay function produced a
non-empty trajectory that could be rendered without geometric degeneracy and
exceeded a minimal admissible length (empirically $> 20\,\mu\mathrm{m}$) with
sufficient transition continuity along the $k$NN graph.

For each patch and mode $m \in \{\mathrm{med},\mathrm{conv}\}$, we defined
the path drawing rate (PDR) as
\[
\mathrm{PDR}^{(m)}_{\mathrm{real}}
  = \frac{N^{(m)}_{\mathrm{real,\,drawn}}}
         {N^{(m)}_{\mathrm{real,\,total}}},\qquad
\mathrm{PDR}^{(m)}_{\mathrm{null}}
  = \frac{N^{(m)}_{\mathrm{null,\,drawn}}}
         {N^{(m)}_{\mathrm{null,\,total}}}.
\]
These patch-wise path drawing rates provide an empirical measure of how
readily triplets give rise to coherent spatial flow patterns under the
observed tissue architecture. To statistically evaluate whether such spatial
coherence occurred more frequently for SpeF--Phixer triplets than expected
under random gene assignment, we compared
$\mathrm{PDR}^{(m)}_{\mathrm{real}}$ and $\mathrm{PDR}^{(m)}_{\mathrm{null}}$
using patch-level $2{\times}2$ contingency tables and one-sided Fisher's exact
tests, reporting odds ratios and exact $p$-values. This null benchmarking
procedure quantifies the extent to which the SpeF--Phixer consensus triplets
preferentially correspond to spatially realized gene regulatory flows.

\subsubsection*{Database-backed direction validation}
To evaluate whether the inferred directional relations ($X\to Y$, $Y\to Z$, 
and $X\to Z$) in real triplets were consistent with known gene regulatory 
interactions, we compared each ordered gene pair against four curated 
databases: RegNetwork (TF--target), CytoSig, CytReg (TF--cytokine 
regulation), and Reactome (protein--protein interactions). All databases 
were harmonized to upper-case HGNC gene symbols and merged into a unified 
reference set of directed edges.

For each triplet $(X,Y,Z)$, we assessed the presence of the ordered pairs 
$X\to Y$, $Y\to X$, $Y\to Z$, $Z\to Y$, $X\to Z$, and $Z\to X$ in the 
reference set. An edge was counted as ``database-backed'' if it appeared 
in any of the four sources. Directional precision metrics were then 
computed separately for mediated triplets and converged triplets, enabling 
evaluation of whether SpeF--Phixer preferentially inferred the biologically 
correct direction for each relationship.

Because database coverage is sparse and non-uniform, the evaluation does 
not aim to capture all true regulatory interactions, but instead assesses 
whether the distribution of database-backed edges shows a statistically 
meaningful preference for the inferred direction (e.g., $Y\to Z$ vs.\ $Z\to Y$).

\subsection{Implementation details}
All analyses were implemented in Python (v3.8) using NumPy, SciPy, pandas,
scikit-learn, matplotlib, and PyTorch (v1.7).  
Spatial processing, $\phi$-mixing computation, and all bootstrap procedures 
were parallelized using Python's \texttt{multiprocessing} module, with random 
seeds fixed across workers to ensure reproducibility.  
The $\phi$-mixing implementation follows \citet{singh2019inferring}, with 
modifications for spatially extended variables and a common discretization 
scheme applied to all genes within each patch.

Computations were performed on a workstation equipped with an NVIDIA RTX~3060 
GPU and 64\,GB RAM.  
A full SpeF--Phixer run on a single patch---including neighborhood screening, 
signed $\phi$-mixing, and $B{=}300$ bootstrap iterations—typically required 
10--25 minutes on this workstation (2--3 hours on a standard laptop), depending 
on the number of mapped cells and markers.

The complete inference and visualization pipeline is publicly available as 
\textbf{SpeF--Phixer} (v1.0.0) under an open-source license~\citep{spef_phixer}, 
including Jupyter notebooks, preprocessing scripts, and all modules required 
to reproduce the analyses.

\section{Results}

\subsection*{Extraction of stable regulatory structures across tissue patches}
Across the 24 eligible patches, the number of candidate gene triplets was
progressively reduced through the SpeF--Phixer bootstrap--consensus
inference pipeline. Each patch initially contained 1,171 candidate markers
(predefined marker panel plus highly variable genes identified from the
scRNA-seq dataset). A small subset of markers (typically 16 ORF-like or
lowly expressed loci), such as \texttt{C11orf53}, \texttt{C16orf89}, and
\texttt{C2orf42}, did not appear in the mapped single-cell expression table
for that patch and were therefore excluded. This yielded a consistent set of
1,155 \emph{expressed markers after filtering}, which were subsequently used
for constructing spatial gene fields and computing $\phi$-mixing coefficients.
Using these 1,155 patch-specific expressed markers, the $k$NN-based
neighborhood graph contained an average of $3795 \pm 40$ directed neighbor
pairs. This combinatorial structure gave rise to approximately
$3.6{\times}10^{4}$ possible triplets per patch
(mean $36{,}080 \pm 359$).
After bootstrap consensus filtering ($\pi \ge 0.60$), the total number of
retained triplets was reduced by nearly two orders of magnitude, yielding
a mean of $372.6 \pm 189.8$ consensus triplets per patch. The corresponding
consensus edge set contained $449.4 \pm 112.4$ directed edges with
bootstrap support. Collectively, these results show that the bootstrap
aggregation step substantially narrows the high-dimensional search space
while isolating a reproducible subset of regulatory relationships that are
consistently recovered across spatially resampled gene--cell configurations.
These summary statistics across patches are listed in
Table~\ref{tab:patch_pipeline_summary}.

\begin{table}[htbp]
\centering
\caption{Summary of gene- and triplet-level quantities across 24 eligible patches. 
Values represent mean $\pm$ SD (range in parentheses).}
\label{tab:patch_pipeline_summary}
\adjustbox{width=\columnwidth}{		
\begin{tabular}{lccc}
\hline
Stage & Quantity & Mean $\pm$ SD & Min--Max \\
\hline
Initial marker set & Candidate markers & 1171.0 & (1171--1171) \\
After expression filtering & Expressed markers & 1155.0 & (1155--1155) \\
Neighborhood graph & Directed neighbor pairs & $3795 \pm 40$ & (3709--3860) \\
Triplet candidates & Raw triplets & $36080 \pm 359$ & (35334--36670) \\
Bootstrap evaluation & Evaluated triplets & 8000.0 & (8000--8000) \\
Consensus stage & Consensus triplets & $372.6 \pm 189.8$ & (92--690) \\
Consensus edges & Directed edges & $449.4 \pm 112.4$ & (259--658) \\
\hline
\end{tabular}
}	
\end{table}

\subsection*{Spatial validation of inferred causal flows through real–null comparison}

Across eligible patches, SpeF--Phixer triplets exhibited substantially higher
path drawing rates than randomly sampled null triplets.
For mediated-mode tracing, real triplets produced spatial bundles in
354/1123 cases (31.5\%), whereas null triplets succeeded in only 155/1123 cases
(13.8\%), corresponding to an odds ratio of 2.28 and a two-sided Fisher's
exact test $p < 0.0001$. For convergent-mode tracing, real triplets yielded
1025/3187 successful bundles (32.2\%) compared to 146/3187 (4.6\%) for null
triplets, giving an odds ratio of 7.02 (two-sided Fisher's exact test
$p < 0.0001$). These results indicate that the causal chains inferred by
SpeF--Phixer are spatially instantiated at rates far exceeding those expected
under random gene assignments, demonstrating that the identified regulatory
flows correspond to non-random, spatially coherent signaling structures in
the tumor microenvironment.

\begin{table}[htbp]
\centering
\caption{Fisher's exact test for path drawing rate (mediated mode).}
\adjustbox{width=\columnwidth}{		
\begin{tabular}{lcc}
\hline
& Successfully drawn & Not drawn \\
\hline
Real triplets & 354 & 769 \\
Null triplets & 155 & 968 \\
\hline
Odds ratio & 2.28 \\
Two-sided Fisher's exact $p$-value & $< 0.0001$ \\
\hline
\end{tabular}
}	
\end{table}

\begin{table}[htbp]
\centering
\caption{Fisher's exact test for path drawing rate (converged mode).}
\adjustbox{width=\columnwidth}{		
\begin{tabular}{lcc}
\hline
& Successfully drawn & Not drawn \\
\hline
Real triplets & 1025 & 2162 \\
Null triplets & 146 & 3041 \\
\hline
Odds ratio & 7.02 \\
Two-sided Fisher's exact $p$-value & $< 0.0001$ \\
\hline
\end{tabular}
}	
\end{table}

\subsection*{Direction validation against regulatory databases}
To examine whether the inferred triplet directions were concordant with
established gene regulatory relationships, we compared each directed edge
with curated databases (RegNetwork, CytoSig, CytReg, Reactome). Because
coverage is limited and heterogeneous across these resources, the goal was
not to obtain an absolute accuracy estimate but to test for directional
asymmetry when database evidence was available.

For mediated triplets (1{,}123 total), overall database coverage was low.
At the upstream edge (X$\to$Y vs.\ Y$\to$X), only 32 edges had database
support (13 vs.\ 19), yielding no clear directional preference (40.6\%
precision). At the downstream edge (Y$\to$Z vs.\ Z$\to$Y), 76 edges were
backed (47 vs.\ 29), corresponding to 61.8\% precision for the inferred
Y$\to$Z direction (two-sided Fisher's exact test $p = 0.023$), indicating
modest but statistically significant directional bias consistent with the
hypothesized causal flow.

In contrast, converged triplets (3{,}187 total) showed a reproducible asymmetry
at the downstream edge: Y$\to$Z matches were more frequent than Z$\to$Y
(158 vs.\ 86 out of 244 database-backed edges), corresponding to 64.8\%
precision for the inferred direction (one-sided Fisher's exact test
$p < 0.0001$). The direct convergent edge (X$\to$Z vs.\ Z$\to$X) showed
no directional preference (51 vs.\ 67 out of 118 backed edges, 43.2\%
precision, $p = 0.94$), suggesting that this edge may reflect correlational
rather than strictly directional regulatory relationships. Notably, no
upstream edges (X$\to$Y or Y$\to$X) were found in the databases for
converged mode, likely reflecting the sparse coverage of such regulatory
patterns in current knowledge bases.

Taken together with the higher spatial path drawing rate for real versus
null triplets, these results justify using the Y$\to$Z downstream edge as
the backbone when assembling larger-scale SpeF--Phixer gene regulatory
networks, as this component shows both spatial coherence and directional
consistency with established biological knowledge.

\subsection*{Spatial visualization of inferred causal flows}

\begin{figure}[htbp]
\centering
\begin{subfigure}[b]{0.48\columnwidth}
    \centering
    \includegraphics[width=\columnwidth]{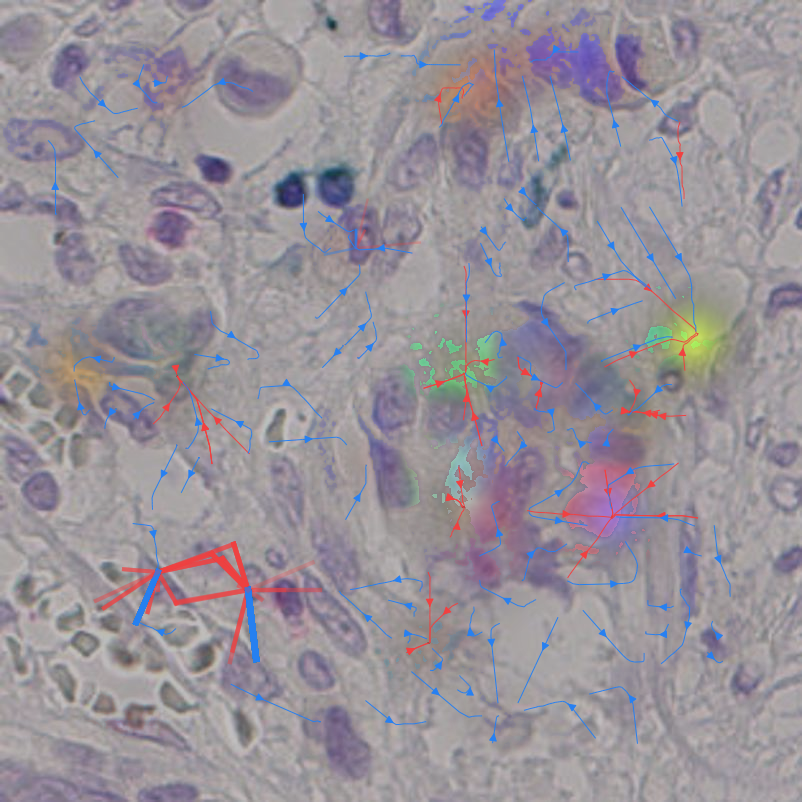}
    \caption{Mediated triplet (REG3A $\rightarrow$ LCN15 $\rightarrow$ FABP1). 
    Red: $X$ expression (REG3A), Green: $Y$ expression (LCN15), 
    Blue: $Z$ expression (FABP1). 
    Red arrows: $X\to Y$ path; blue arrows: $Y\to Z$ path.
    Streamlines follow the same color convention.}
    \label{fig:overlay_mediation}
\end{subfigure}
\hfill
\begin{subfigure}[b]{0.48\columnwidth}
    \centering
    \includegraphics[width=\columnwidth]{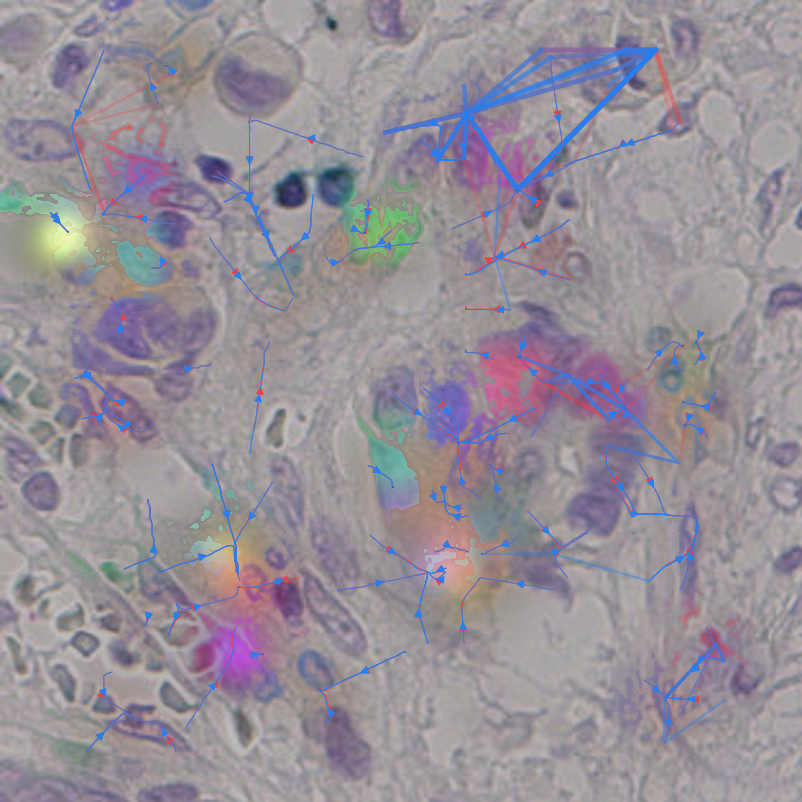}
    \caption{Convergent triplet (NRG1, MMP9 $\rightarrow$ HLA-DRA). 
    Red: $X$ expression (NRG1), Green: $Y$ expression (MMP9), 
    Blue: $Z$ expression (HLA-DRA). 
    Red arrows: $X\to Z$ path; blue arrows: $Y\to Z$ path.
    Streamlines follow the same color convention.}
    \label{fig:overlay_converged}
\end{subfigure}
\caption{
Spatial overlay of SpeF--Phixer flow fields for representative mediated and convergent triplets.
For all panels, color channels represent gene expression (Red: $X$, Green: $Y$, Blue: $Z$),
and arrow colors indicate the inferred causal direction for each edge.}
\label{fig:overlay_example}
\end{figure}

Spatial flow patterns derived from the KDE edge-bundling and streamline
modules revealed characteristic geometric signatures for mediated and
convergent triplets.  
Representative examples are shown in
Figure \ref{fig:overlay_example}, corresponding to the same visualization
procedures described in the Methods section.

In the \emph{mediated} triplet
(REG3A $\rightarrow$ LCN15 $\rightarrow$ FABP1)
(Figure \ref{fig:overlay_mediation}),
red bundles and streamlines denote the upstream edge
$X\!\rightarrow\!Y$ (REG3A $\rightarrow$ LCN15),
and blue bundles and streamlines denote the downstream edge
$Y\!\rightarrow\!Z$ (LCN15 $\rightarrow$ FABP1).
Both modalities consistently aligned to form a coherent two-step spatial
flow, with red trajectories converging into blue trajectories.
Gene-expression overlays followed the pre-defined RGB order
(X: red, Y: green, Z: blue),
highlighting the spatial continuity of the inferred causal chain across
adjacent histological structures.

In contrast, the \emph{convergent} triplet
(NRG1, MMP9 $\rightarrow$ HLA-DRA)
(Figure \ref{fig:overlay_converged})
exhibited a distinct geometric configuration.
Red bundles/streamlines correspond to the $X\!\rightarrow\!Z$ flow
(NRG1 $\rightarrow$ HLA\textrm{-}DRA), and blue bundles/streamlines to the
$Y\!\rightarrow\!Z$ flow (MMP9 $\rightarrow$ HLA\textrm{-}DRA),
both converging onto Z from spatially separated upstream regions.
The RGB expression overlays (NRG1: red, MMP9: green, HLA-DRA: blue)
revealed spatially distinct upstream gene fields feeding into a common
downstream region, matching the geometry expected for a convergent causal
structure.

\begin{figure}[htbp]
\centering
\begin{subfigure}[b]{0.48\columnwidth}
    \centering
    \includegraphics[width=\columnwidth]{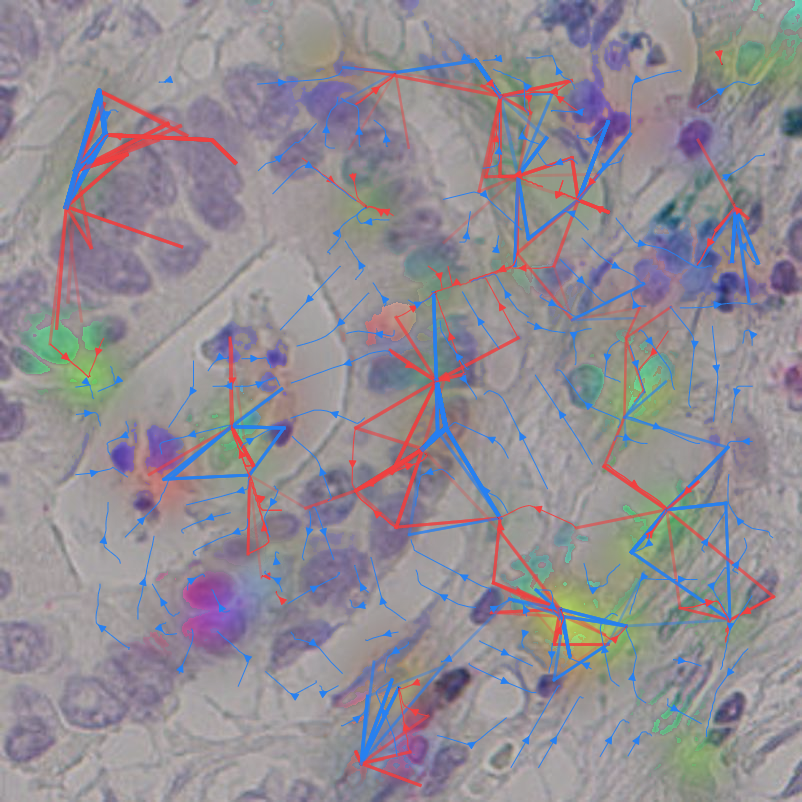}
    \caption{Additional mediated example (CCR6 $\rightarrow$ PDGFRB $\rightarrow$ HIF1A).
    Red indicates the $X\!\rightarrow Y$ flow and blue the $Y\!\rightarrow Z$ flow.
    Gene expression channels: $X=$Red, $Y=$Green, $Z=$Blue.}
    \label{fig:overlay_mediation_ccr6}
\end{subfigure}
\hfill
\begin{subfigure}[b]{0.48\columnwidth}
    \centering
    \includegraphics[width=\columnwidth]{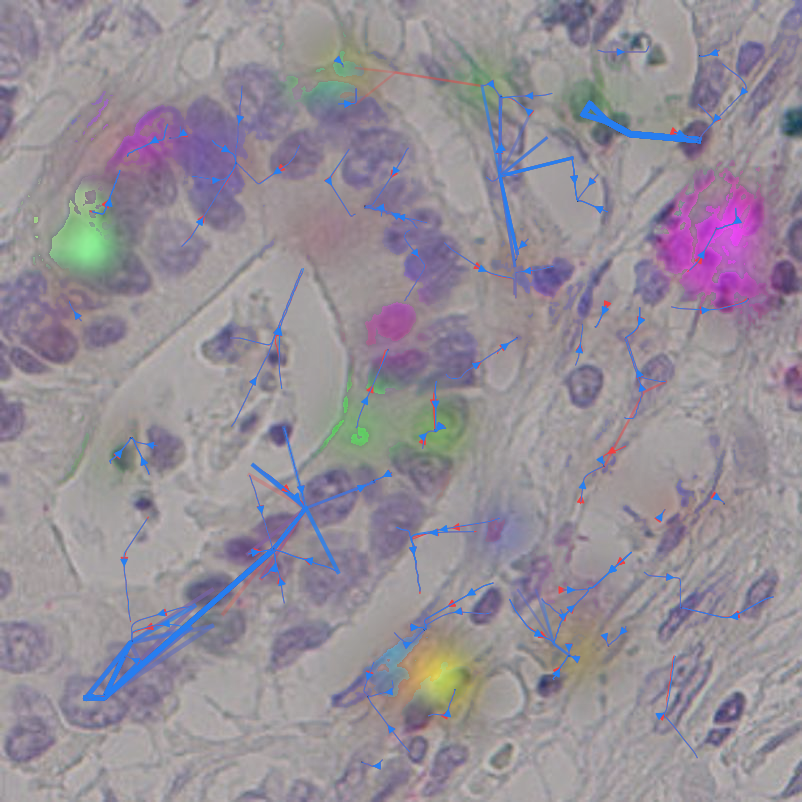}
    \caption{Additional converged example (NCAM1 $\rightarrow$ ITGAE and GNLY $\rightarrow$ ITGAE). 
    Red shows the $X\!\rightarrow Z$ path and blue the $Y\!\rightarrow Z$ path.
    Gene expression channels: $X=$Red, $Y=$Green, $Z=$Blue.}
    \label{fig:overlay_converged_ncam}
\end{subfigure}
\caption{
Examples of mediated and converged causal triplets inferred by SpeF--Phixer.
Each image overlays (i) spatial KDE-bundled causal paths and (ii) gene expression
streamlines for triplet members. Red and blue indicate upstream and downstream
directions depending on mediation or convergence mode. Gene expression images
use RGB encoding (X: red, Y: green, Z: blue).}
\label{fig:overlay_example_extended}
\end{figure}

In addition, spatial flow analyses of an independent patch illustrated
further examples of both mediated and convergent configurations.
The mediated triplet
(CCR6 $\rightarrow$ PDGFRB $\rightarrow$ HIF1A)
(Figure \ref{fig:overlay_mediation_ccr6})
displayed a clear two-step trajectory in which red bundles/streamlines
(CCR6 $\rightarrow$ PDGFRB) smoothly connected into blue bundles/streamlines
(PDGFRB $\rightarrow$ HIF1A).
The RGB overlays (CCR6: red, PDGFRB: green, HIF1A: blue)
again highlighted a continuous transition in spatial gene expression
consistent with a mediating regulatory sequence.

By contrast, the convergent triplet
(NCAM1, GNLY $\rightarrow$ ITGAE)
(Figure \ref{fig:overlay_converged_ncam})
showed two spatially distinct upstream flows.
Red bundles/streamlines (NCAM1 $\rightarrow$ ITGAE) and
blue bundles/streamlines (GNLY $\rightarrow$ ITGAE)
approached the downstream gene from separate regions of the tissue.
The RGB overlays (NCAM1: red, GNLY: green, ITGAE: blue)
revealed distinct expression domains funneled into the same ITGAE-positive
area, consistent with a convergent regulatory configuration.

Across patches, these qualitative patterns---serial alignment of upstream and
downstream segments in mediated triplets versus spatially separated inflows
in convergent triplets---were consistently reproduced in both the bundled-path
and streamline layers. Rather than encoding explicit direction in the bundles
themselves, the coincidence of high-density pathway bundles with regions of
strong expression gradients for the corresponding genes provides spatial
support for the two distinct causal configurations inferred by the
SpeF--Phixer pipeline.

\section{Discussion}

\subsection*{Interpreting spatially resolved directed dependencies}
SpeF--Phixer integrates histological spatial structure with scRNA-seq--derived
gene expression to infer directed dependencies among genes on a per-patch
basis. While pairwise $\phi$-mixing coefficients quantify asymmetric
dependencies under non-Gaussian and non-linear conditions, their interpretation
is enhanced by embedding them within the spatial context of the tissue.
Because $\phi(X_j \mid X_i)$ measures directional divergence of conditional
distributions, its magnitude indicates regulatory influence but does not imply
the existence of a direct causal mechanism. The spatially extended formulation
used here, based on diffused gene fields, ensures that local microstructural
heterogeneity---particularly tumor–stroma interfaces---contributes to the
effective dependency landscape. These directed dependencies therefore represent
spatially contextualized regulatory tendencies rather than purely
correlation-based associations.

\subsection*{Mediation and convergence as higher-order causal structures}
Although pairwise $\Phi$-mixing relations identify promising directed edges,
they cannot distinguish mediated chains from convergent structures. Both
patterns may generate similar pairwise dependencies
($\phi_{Y|X}$, $\phi_{Z|Y}$, $\phi_{Z|X}$), and thus the $\Phi$-matrix alone
cannot assess conditional independence within triplets. Incorporating
pseudotime-adjusted generalized additive models (GAMs) enables conditional
testing of higher-order causal structures. In a mediated chain, the effect of
$X$ on $Z$ should diminish once $Y$ is included, whereas in a convergent
structure both $X$ and $Y$ retain significant partial effects on $Z$. Repeated
GAM fitting under block bootstrap resampling stabilizes these assignments and
provides support scores for mediation and convergence. This higher-order
dissection is essential because it identifies regulatory motifs not detectable
from pairwise analysis alone, revealing whether spatial causal flow is routed
through intermediate genes or arises from convergent upstream inputs.

\subsection*{Spatial validation through real--null benchmarking}
A central question is whether the spatial flow patterns inferred from
SpeF--Phixer triplets simply reflect generic properties of the tissue
architecture or whether they represent genuine biological structure.
The real--null comparison directly addresses this point by evaluating the
path drawing rate (PDR) for real versus randomly assembled triplets under
identical spatial and expression distributions. The markedly higher PDR for
real triplets---with odds ratios of 2.28 for mediated-mode and 7.02 for
convergent-mode tracing---demonstrates that the inferred triplets are far more
likely to generate spatially coherent flow patterns than expected under random
gene assignment. These results indicate that the directed structures identified
by SpeF--Phixer correspond to non-random, spatially instantiated signaling
flows embedded within the tumor microenvironment. The strong real–null
separation further supports the validity of using spatial coherence as an
orthogonal criterion for evaluating regulatory plausibility.

\subsection*{Directional validation against curated regulatory databases}
Although database coverage is sparse and incomplete, comparing SpeF--Phixer
edges with curated regulatory resources (RegNetwork, CytoSig, CytReg, Reactome)
provides an independent assessment of directional plausibility. For mediated
triplets, downstream edges $Y\!\to\!Z$ exhibited a statistically significant
directional preference compared with $Z\!\to\!Y$ (61.8\% precision, $p=0.023$).
For convergent triplets, $Y\!\to\!Z$ was strongly favored over $Z\!\to\!Y$
(64.8\% precision, $p<0.0001$). These findings indicate that the inferred
downstream edges align more frequently with existing biological knowledge than
expected by chance, despite low database coverage. In contrast, direct
convergent edges ($X\!\to\!Z$ vs.\ $Z\!\to\!X$) showed no directional bias,
suggesting that these edges may capture co-regulation or shared upstream
drivers rather than strict causal directions. Together with the spatial
validation, these results support using the downstream $Y\!\to\!Z$ relation as
the most reliable component when assembling larger-scale regulatory networks.

\subsection*{Integration of spatial flow with biological interpretation}
The KDE-based edge bundling and streamline visualization provide
two complementary views of regulatory flow: trajectory density (bundles) and
expression-gradient direction (streamlines). Their conjunction highlights
regions where both flow representations agree, revealing robust spatial
hotspots of regulatory activity. Notably, many high-confidence triplets occur
near tumor–stroma boundaries, where gradients of immune and epithelial states
intersect. This spatial convergence is consistent with known biology: immune
activation, stromal remodeling, and epithelial stress signaling often interact
through spatially constrained channels. SpeF--Phixer enables direct
visualization of such region-specific regulatory flows, offering an interpretable
bridge between molecular measurements and microenvironmental structure.

\subsection*{Methodological limitations and future extensions}

Several methodological considerations should be noted.  
First, although SpeF--Phixer integrates histological structure with 
single-cell gene expression, the scRNA-seq profiles used in this study 
were derived from publicly available datasets and are not patient-matched 
to the specific WSI specimens analyzed. Consequently, the mapped gene 
expression fields represent a probabilistic reconstruction rather than 
a true one-to-one molecular readout from the same physical tissue.  
Nevertheless, for the purpose of concept validation, this design provides 
a rigorous test of whether spatially projected gene expression---even when 
obtained from an external cohort---can support consistent inference of 
directional regulatory triplets. The reproducibility of consensus 
triplets across patches, together with their superior spatial path drawing 
rates compared with null structures, indicates that the approach remains 
robust even under this non-ideal but practically relevant scenario.

Second, the current mapping procedure does not capture gene–protein 
discordance or context-dependent transcriptional dynamics that may arise 
locally within tumor or stromal niches. Future work incorporating 
matched multi-omic profiling or antibody-based spatial readouts may 
further improve the biological fidelity of the projected gene fields.

Third, although spatial transcriptomics (ST) platforms offer direct 
molecular measurements on tissue sections, current ST technologies are 
limited by coarse spatial resolution, restricted cell-type specificity, 
and inability to repeatedly resample spatial configurations.  
In contrast, SpeF--Phixer leverages the fine spatial resolution of WSI 
and enables \emph{bootstrap-based simulation} across thousands of virtual 
resampled spatial environments. This repeated resampling---not feasible 
with ST---allows estimation of consensus triplets, stability metrics, and 
directional confidence scores, providing a statistical characterization of 
spatial regulatory structure that extends beyond single-shot spatial assays.

Finally, the supervised steps of the pipeline, including neighborhood 
screening and discretization, introduce tunable hyperparameters.  
While benchmarks across patches showed stability, adaptive parameter 
selection or Bayesian hierarchical models may further improve robustness.  
Extending SpeF--Phixer to incorporate continuous-valued dependency 
measures, dynamic pseudotime warping, or joint modeling of multiple 
tissue regions could broaden its applicability to diverse spatial 
transcriptomic and histologic scenarios.

\section{Conclusion}

We developed SpeF--Phixer, a spatially aware extension of the 
$\phi$-mixing framework that integrates histology-derived cell maps with 
mapped scRNA-seq expression fields to infer directed gene relationships.  
By combining neighborhood screening, signed $\phi$-mixing, and 
bootstrap consensus, the method identifies reproducible three-gene 
regulatory structures that remain coherent across spatial resampling.

Directional validation showed that downstream edges ($Y\!\rightarrow Z$) of 
consensus triplets exhibit statistically significant agreement with curated 
regulatory databases, supporting their reliability for network assembly.  
These results demonstrate that SpeF--Phixer can recover spatially coherent 
and directionally consistent regulatory flows, enabling construction of 
interpretable causal gene networks directly from histological tissue data.

\section*{Acknowledgements}

We would 
like to express our sincere gratitude to the following individuals for their valuable contributions to the precise image tagging: To the graduate students and technical assistants from Kobe University who assisted with the annotation tasks: Abe T, Adachi Y, Agawa K, Ando M, Fukuda S, Imai M, Ito R, Kagiyama H, Konaka R, Miyake T, Mukoyama T, Okazoe Y, Tachibana T, Takahashi T, Ueda Y and Yasuda K. We also extend our appreciation to the registered annotators who were engaged by the AMAIC: Adachi K, Akima J, Aoki S, Ichikawa K, Kanto T, Kawase Y, Kimura M, Miura R, Sirasawa H, Sotani K and Yuki A. Their professional and diligent efforts greatly enhanced the quality of the dataset.
This study is supported by the Grants-in-Aid for Scientific Research from the Ministry of Education, Culture, Sports, Science and Technology of Japan (MEXT; 24K10381 to TN, 23K08171 to KY, and 21K09167 to MF).

\section*{Declaration of generative AI and AI-assisted technologies in the writing process}

During the preparation of this work the authors used Claude Opus 4.1 (Anthropic) 
and ChatGPT 5.1 (OpenAI) to check for errors in analysis software code and to 
improve the language and readability of the manuscript text. After using these 
tools, the authors reviewed and edited the content as needed and take full 
responsibility for the content of the publication.

\printcredits

\bibliographystyle{cas-model2-names}

\bibliography{ref}



\end{document}